# Software Sustainability: A Design Case for Achieving Sustainable Pension Services in a Developing Country.

Mikhail Ola Adisa
LUT School of Engineering (LENS)
LUT University, Finland
mikhail.adisa@lut.fi

Shola Oyedeji
LUT School of Engineering (LENS)
LUT University, Finland
shola.oyedeji @lut.fi

Jari Porras
LUT School of Engineering (LENS)
LUT University, Aalto University, Finland
jari.porras@lut.fi

*Abstract*— The need for efficient and sustainable software to improve business and achieve goals cannot be over-emphasized. Sustainable digital services and product delivery cannot be achieved without embracing sustainable software design practices. Despite the current research progress on software sustainability, most software development practitioners in developing countries are unclear about what constitutes software sustainability and often lack the proper understanding of how to implement it in their specific industry domain. Research efforts from software engineering focused on promoting software sustainability awareness in developed countries, and fewer efforts have been channeled to studying the same awareness in developing countries. This has affected the level of awareness about sustainable software design practices in most developing countries. This research investigates the awareness of software sustainability in the Nigerian pension industry and its challenges among practitioners. The software development practitioners were engaged and interviewed. We offered ways to mitigate the identified challenges and promote the awareness of software sustainability in the pension industry. Our findings further show that, with the right sustainability knowledge, the software practitioners in the pension industry have the potential to support their organization's sustainable culture and improve the efficiency of product design and service delivery.

## I. Introduction

The development and use of sustainable software systems can bring positive environmental impacts by optimizing service delivery through a revamped business process and reducing the energy and resource requirements of the final digital products. The desire to achieve sustainable development is a driving force in software development to bring new opportunities, cut costs, add value, and gain a competitive advantage [1], [2]. Sustainable software can be interpreted in two ways: [3] the software code being sustainable, agnostic of purpose, or the software purpose supporting sustainability goals. In addition, sustainable software can be viewed as energy-efficient that [1] minimizes the environmental impact of the processes it supports and positively impacts social and economic sustainability. These impacts can occur directly (energy), indirectly (mitigated by service), or as rebound effects [4].

Furthermore, the software development process is a major driving force toward the pension industry's innovation pursuit to deliver efficient services and launch new digital products that positively impact the environment. However, the current practices and low awareness of sustainable software design practices among software development practitioners in this industry calls for concern. The Nigerian pension industry is under the National Pension Commission's (PenCom) supervision as a contributory pension scheme for employees in the private and public sectors. The current value of this industry stood at N13trillion (~$31.69 billion), resulting from a total contribution of 9.46m clients spreading across 22 licensed Pension Fund Administrators (PFAs) [5]–[7]. As a result, the Nigerian pension industry serves as a starting point for investigating challenges of sustainability awareness and sustainable software design practices in developing countries.

## II. Aim and Objectives

This research investigates the software development practices among software development practitioners in the Nigerian pension industry to understand their awareness of sustainable software development and the critical challenges of delivering products and services that meet the industry's goal and engage clients sustainably while considering the environmental impact of the organization's activities.

The research questions below were designed to achieve the aim and objectives of the research:

RQ 1: *What are the current awareness and practices of sustainable software design in the Nigerian pension industry?*

RQ 2: *What challenges are identified by key software development practitioners for embracing sustainable design practice?*

RQ 3: *How could SusAf framework be used to increase sustainability awareness and address the challenges in RQ 2?*

## III. Methodology

We employed a qualitative research methodology by conducting semi-structured interview sessions online via zoom meetings for the available software practitioners. The respondents were selected from the registered PFAs in Nigeria and had between two to over ten years of working experience in the industry's software development-related processes. In addition, two of the respondents are currently the head of their respective ICT departments and have over ten years of experience, respectively. In general, we were limited by the availability and willingness of all the targeted respondents to participate as at the time of conducting the interview sessions; hence 15 respondents were eventually considered for the final interview session.

The data collection and analysis were guided by using the Grounded Theory (GT) [8][9][10] research method, following the interview sessions conducted for the selected respondents to have a clear overview of their understanding of sustainability and what sustainability means in the software development process for their industry. We decided to use GT due to its systematic approach to generating theory from data [8] and the opportunity to identify new perspectives and theories from established academic areas [9], such as software engineering. Furthermore, this research identifies the respondents' perspectives (using collected data) about software sustainability. The GT, therefore, offers a suitable way for data gathering, analysis, and extracting the proper context from a complex scenario, allowing researchers to have a clearer understanding of a specific substantive issue [10].

## IV. FINDINGS

This research provides a clear insight into the current perception of software sustainability among software practitioners in the Nigerian pension industry and how we can further engage them to embrace sustainable design practice.

In answering the RQ 1: *What are the current awareness and practices of sustainable software design in the Nigerian pension industry?* The current awareness about software sustainability is inadequate among the practitioners. There is a wrong perception that sustainability in ICT only applies to the hardware infrastructure and "*…sustainable software as focusing on delivering usability*". Furthermore, at no point was sustainability considered as a quality attribute of software in their design process.

For the RQ 2: *What challenges are identified by key software development practitioners for embracing sustainable design practice?* The identified key challenges are (i) the fear of altering their current software design approach and (ii) The epileptic electricity situation in the country, which forced all the PFAs to rely on fossil-powered generators for their Datacenters in addition to inverter systems. In their own words, "*The emissions from the PFAs like other industries in the country will continue to increase if the business must continue*". In the past eight years, the country has averaged 4,500 megawatts of electricity supply, leading to incessant blackouts and (iii) The bureaucracy surrounding software development practitioners' training on sustainable software design process and procurement of relevant tools often delayed the management approval.

With the RQ 3: *How could the SusAf framework be used to increase sustainability awareness and address the challenges in RQ 2?* There is insufficient knowledge of any suitable sustainability framework among key stakeholders in the industry. However, by engaging the software development practitioners with the guidelines from the Sustainability awareness framework (SusAf) question sheets [12], we asserted that the practitioners' transition to a sustainable software design process could be achieved by adopting the SusAf in their software development practices.

The SusAf framework helps to highlight the possible impacts of their activities, solidify their knowledge, promote sustainability awareness, and empower them to embrace the sustainable software design process at every phase of their software development life cycle. For example, a one-stop portal accessible to the client could reduce physical visitation to the office, less paper, and less air pollution on the road due to less traveling. In the same way, increased use of fossil power on-premise data centers can increase $CO_2$ emissions from the industry to the environment. A further demonstration of the potential chain of effects is shown in Figure 1 below. It is ongoing work as we plan to extend the research to cover more software development practitioners in other industries.

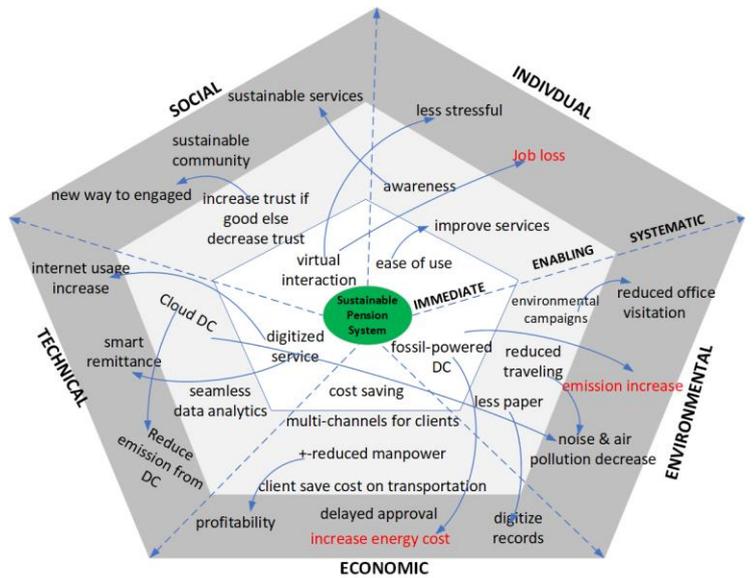

Fig. 1. The Sustainability Analysis Radar Chart

## V. DISCUSSION AND CONCLUSIONS

This research serves as the stepping stone for further research into the ICT sustainability activities and challenges in developing countries. We used the Nigeria pension industry as our starting point to understand the challenges of sustainable software design among software development practitioners in developing countries. Our findings align with other research work [12], [13] that the potential impacts of software systems can be managed effectively when software development practitioners take responsibility for the sustainability impacts and effects of their software products and services. In addition, the well-established pension industry offers a better ground for initiating awareness of software sustainability among the industry's software development practitioners.

The SusAf framework [12] provides simple and accessible ways to elicit awareness among the stakeholders about the impacts software systems could have on sustainability. Furthermore, the SusAf framework offers us the guidelines for identifying the chain of effects in the actual software development processes. As a result, we hope to empower the pension industry to deliver digital pension products and services with sustainability at the core of their design strategy.